\documentclass[10pt]{article}
\usepackage{epsfig,psfrag,amsmath,amssymb,latexsym}
\usepackage{amscd}
\pdfoutput=1
\usepackage{float}
\usepackage{subfig}
\usepackage{amssymb, graphicx, multicol, color, epic, eepicemu, epsfig, float, enumerate}
\usepackage{lscape}
\usepackage{amsthm}
\usepackage{natbib}
\usepackage{hyperref}

\usepackage[english]{babel}
\usepackage{lmodern}

\def\X{{\bf X}}
\def\x{{\bf x}}
\def\university#1{{\sl \begin{center} #1 \vspace{5pt} \end{center} } }
\def\inst#1{\vspace{1pt} \unskip$^{#1}$}
\begin{document}
%
\title{Comparison of hidden Markov chain models and hidden Markov random field models in estimation of computed tomography images}

\author{Kristi Kuljus\inst{1}\footnote{Corresponding author, e-mail: kristi.kuljus@ut.ee}, Fekadu L.~Bayisa\inst{2}, David Bolin\inst{3}, J\"uri Lember\inst{1} and Jun Yu\inst{2}}
%
\date{April 17, 2017}
\maketitle
\university{\inst{1}University of Tartu, Estonia; \inst{2}Ume{\aa} University, Sweden; \\
\inst{3}Chalmers University of Technology and University of Gothenburg, Sweden }
%
\begin{abstract}
There is an interest to replace computed tomography (CT) images with magnetic resonance (MR) images for a number of diagnostic and therapeutic workflows. In this article, predicting CT images from a number of
magnetic resonance imaging (MRI) sequences using regression approach is explored. Two principal areas of application for estimated CT images are dose calculations in MRI based radiotherapy treatment planning and attenuation correction for positron emission tomography (PET)/MRI.  The main purpose of this work is to investigate the performance of hidden Markov (chain) models (HMMs) in comparison to hidden Markov random field (HMRF) models when predicting CT images of head. Our study shows that HMMs have clear advantages over HMRF models in this particular application. Obtained results suggest that HMMs deserve a further study for investigating their potential in  modeling applications where the most natural theoretical choice would be the class of HMRF models.
\end{abstract}
\section{Introduction} \label{sissejuh}
Magnetic resonance imaging (MRI) and computed tomography (CT) are two different medical imaging technologies that enable to image the anatomy of the human body. Images are widely used for medical diagnostics. The two technologies provide complementary information, both have advantages and limitations. For most MRI sequences, the soft tissue contrast is superior to that of CT. Therefore, MRI provides a considerable advantage over CT when identifying or delineating tumors. Varying the imaging parameters enables to obtain MR images better suited for particular purposes. A major disadvantage of MRI is that the contrast between air and bone is poor. Since bone has low hydrogen density, both bone and air appear dark in an MR image. CT images are acquired using ionizing radiation, which is both an advantage and disadvantage of CT imaging. On the one hand, CT intensity values reflect the radiation interaction properties of the tissues that are needed for dose planning in radiotherapy. On the other hand, being exposed to radiation during CT examinations is associated with a risk to induce cancer, because radiation can damage body cells. Further, CT gives better images of bones and CT investigation takes shorter time. To summarize, one can say that if MRI and CT could yield the same diagnostic information, then MRI would be preferred and recommended over CT.
Therefore, it is desirable to investigate whether one can substitute CT images with so-called pseudo-CT images (sometimes also called CT substitutes) that are estimated from MRI sequences. The question is how good CT estimates are we able to obtain and for what purposes are these pseudo-CT images feasible.

There is a crucial need for estimating CT images from MR images. Two very important application areas for pseudo-CT images are attenuation correction in positron emission tomography (PET)/MRI and MRI-based dose planning in radiotherapy. A comprehensive state-of-the-art overview of MRI-guided attenuation correction methods in PET/MRI can be found in \citet{zaidiAC}. The existing works in the area of deriving CT images from MR images can be roughly divided into four classes (see \citet{forest} and the references therein): tissue segmentation based methods, atlas-based methods, learning-based methods, integration of atlas-based and pattern recognition methods. The latest contribution to deriving pseudo-CT images in the class of learning-based methods is the random forest method \citep{forest}. For a short overview of different pseudo-CT generating methods, we refer also to \citet{adamThesis} and the references therein.

In this article we continue exploring a voxel-wise direct conversion method introduced in \citet{adam1} and further studied in \citet{adam2,adam3}, belonging to the class of learning-based methods. It was seen in these works that the studied regression method provides pseudo-CT images with satisfying quality for dose calculations and attenuation correction. The idea of learning-based methods is that a model for predicting CT images from MR images is estimated using a training data set, and the estimated model is then applied to a target MR image(s). As mentioned above, it is difficult to image bones with MRI, but there is a particular category of MRI sequences with ultrashort echo time (so-called UTE sequences) that can sample the MRI signal from bone before it is lost. Even if with UTE sequences bones can be imaged with weak intensity and poor resolution, these sequences make regression approach for predicting CT images feasible. The main idea is to model the joint distribution of a CT measurement sequence and a number of MRI sequences. The regression function is then obtained as the conditional expectation of CT given the MRI sequences, and pseudo-CT images are derived voxel-wise.  In \citet{adam1,adam2,adam3}, the regression approach was applied with Gaussian mixture models (GMMs), which ignore spatial dependence structure by assuming that voxels are independent. Since obviously voxels in the human body are not spatially independent, it is essential to study how much better substitute images can be derived with more appropriate models, where spatial dependence structure in the data is accounted for.

The main aim of this article is to investigate the performance of hidden Markov (chain) models (HMMs) in comparison to hidden Markov random field (HMRF) models and GMMs, when the purpose is to estimate head CT images from a number of MR sequences. For all the three model classes, the observed variables depend on latent variables indicating what class (i.e.~tissue class or a mixture of tissue classes) the corresponding voxel belongs to. In GMM, the latent variables are assumed to be independent. In the case of first-order neighbourhood structure, every voxel has six neighbours in $\mathbb R^3$. The HMRF model takes into account all the six neighbours, while HMM accounts for two neighbours. Therefore, HMM lies somewhere between GMM and HMRF. Examples in \citet{Fjortoft2003} show that HMRF and HMM can compete in terms of parameter estimation and classification accuracy, while HMM is more robust and computationally much faster and easier to handle. Besides, the classification examples in \citet{Fjortoft2003} demonstrate that classification with HMM represents small structures more precisely, which might be an advantage in our application. The named arguments provide the main motivation for studying HMMs in the problem of deriving CT images from MR images.

The outline of the article is as follows. In Section \ref{datadescr} we describe our data set and how the data is prepared for modeling with HMM. Section \ref{mudelid} explains the parametrizations of HMM and HMRF and describes main model evaluation criteria. In Section \ref{mudvordlus}, the principal modeling results are presented. Section \ref{arutelu} summarizes the main outcomes with a short discussion.
\section{Data description} \label{datadescr}
\begin{figure}[h!]
\begin{center}
{\includegraphics[scale=0.75]{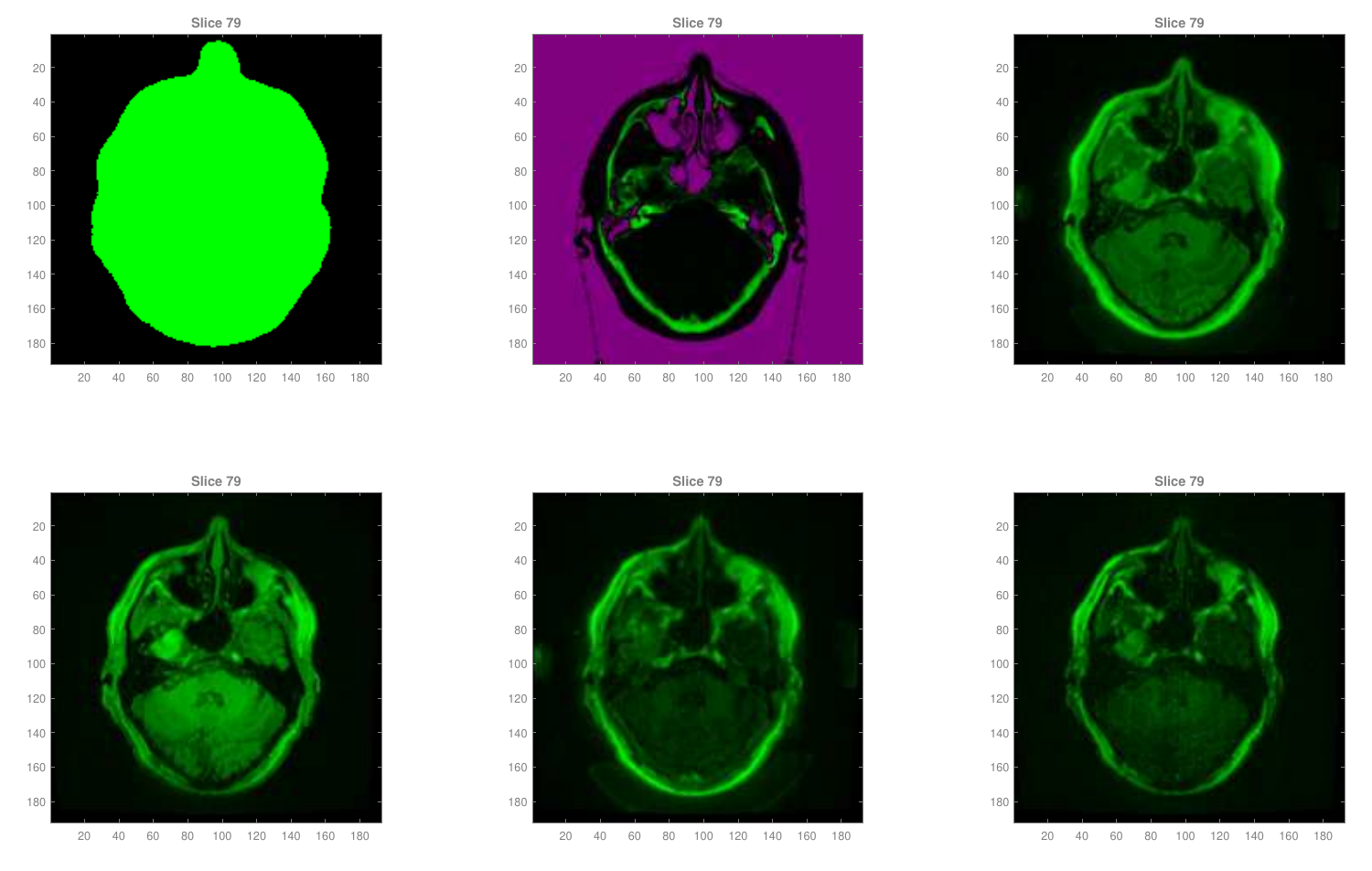}}
\caption{Example of the data: a slice of one head. Row-wise from left to the right: 1) binary mask, 2) CT data, 3)-6) MRI sequences. \label{andmed}}
\end{center}
\end{figure}

We use data from nine patients to evaluate the methods on (five female and four male patients). For each patient, there are measurements on four MRI sequences and one CT image. The MR images were acquired using two dual echo UTE sequences with two different flip angles (10 degrees and 30 degrees). The UTE sequences sampled a first echo (FID) with an echo time of 0.07 ms and a second echo (gradient echo) with an echo time of 3.76 ms. The images were reconstructed to $192\times 192 \times 192$ voxels with a voxel size of 1.33 mm. To achieve voxel-wise correspondence between the images, images of the same patient were co-registered. To separate observation voxels from the air surrounding a head, a binary mask was calculated. Thus, for every voxel we have a $6\times 1$ observation vector with the following variables: 1) binary mask (1 -- observation belongs to the patient, 0 -- surrounding air); 2) CT-value; 3) UTE$_1$-value (70 $\mu s$, $10^\circ$);
4) UTE$_2$-value (3.76 $ms$, $10^\circ$); 5) UTE$_3$-value (70 $\mu s$, $30^\circ$); 6) UTE$_4$-value (3.76 $ms$, $30^\circ$).
Additional details concerning image acquisition, registration and mask calculation can be found in \citet{adam1}. As an example of the data, we have presented in Figure \ref{andmed} a slice of one head for all the six variables.
\subsection{Sequencing data with the Hilbert curve}
A hidden Markov model is a one-dimensional process,
where the observations are assumed to be ordered in space or time.
To be able to apply HMMs to 3-dimensional head data, we have to
`sequence' the data using a space filling curve. A space filling
curve maps 3-dimensional data into a 1-dimensional sequence and
there are several ways to do it. In \citet{sakoglu}, the following space filling curves for ordering 3-dimensional data to 1-dimensional are studied: simple linear ordering (that is row-wise or column-wise ordering), Z-ordering and Hilbert curve ordering. A good space filling curve tries to minimize discontinuity in the head structure, so that anatomically close voxels appear as close as possible also in the corresponding sequence. \citet{sakoglu} show that out of the three studied space filling curves, the Hilbert curve preserves local structure best. After sequencing the data every voxel has two neighbours and thus, we can only make  partial use of the information on a head's spatial structure. One could think of using 3-dimensional HMMs instead (see, for example \citet{3dhmm}), but these 3-dimensional HMMs are quite specific models favouring one certain direction in the space and hence, they lack isotropy.
The Hilbert curve, on the contrary, moves through the space in an isotropic way. Moreover, we aim to keep our HMM as simple as possible in terms of number of parameters and their interpretation, and computational complexity.

In order to perform sequencing with help of the Hilbert curve, we enlarge the data cube of size $192\times 192\times 192$ to a $2^8\times 2^8 \times 2^8$ cube so that the data is in the middle. It follows that when sequencing a head data with the Hilbert curve, we move out of the head and back into the head a number of times. Therefore, we will have to deal with the so called `edge effect'. We have to keep track on the voxels, where we leave the head and where we enter the head. Sometimes these voxels will not be spatial neighbours and we will have a breaking point, that is the sequence is broken into independent segments. When the voxels can be considered as neighbours, we connect them. Thus, after sequencing, the data for one head will consist of a number of independent sequences. For example, for a head of size $1853702$ voxels, we obtain 12239 independent sequences. There are many short sequences (2299 with only one observation, 1884 with two observations etc.), but they correspond to a relatively small amount of voxels.
About $98.8\%$ of the voxels in this head have two neighbours after sequencing.
The maximum sequence length for the considered head is $108205$.
\section{Models for estimating CT images} \label{mudelid}
\subsection{Hidden Markov models}
After mapping the head data with the Hilbert curve, we obtain for each head a number of independent voxel sequences. To the sequenced data we can apply HMMs. Consider an arbitrary voxel sequence of length $n$. Let $Y_t$ denote the value of the CT image for voxel $t$ and let $\X_t=(X_{t,1},\ldots,X_{t,m})$ denote the values of $m$ MR intensities for voxel $t$. We assume that the observed variables
 $(Y_t,\X_t)$ depend on an unobservable or latent variable $Z_t$, which indicates what class the voxel belongs to. In the current application, the classes can be thought of as tissue types or as mixtures of different tissues. An HMM is a double stochastic process $(\{Y,\X \},Z)$, where the observable process $(Y,\X)=\{Y_t,\X_t\}_{t=1}^n$ depends on an unobservable Markov chain $Z=\{Z_t\}_{t=1}^n$. Given $Z$, the variables $(Y,\X)$ are conditionally independent. For any voxel $t$, the hidden Markov chain is in one of the states of $S=\{1,\ldots,K\}$. We assume a first order Markov chain. Thus, for any voxel $t$, a change of state will occur according to a set of probabilities (transition probabilities) associated with the current state as follows:
 \[ P(Z_t=j|Z_{t-1}=i) =p_{ij}, \quad p_{ij}\geq 0 , \quad \sum_j p_{ij} =1, \quad i,j \in S. \]
 Another model component that characterizes HMMs is initial distribution $\pi=(\pi_1,\ldots,\pi_K)$, where $\pi_k=P(Z_1=k)$. Since we move into and out of a head a number of times, the initial distribution can be
 reliably estimated in the current application. We assume that observations for a given state follow a multivariate normal distribution, that is
 \[ (Y_t,\X_t)' \, | \, Z_t=k \, \sim \mathcal{N}(\mu_k,\Sigma_k), \quad k=1,\ldots,K.\]
Modeling the joint distribution with the normal distribution is in agreement with earlier works \citet{adam1,adam2,adam3},
where Gaussian mixture models were applied and thus, the normal distribution
allows a fair comparison. Further, the joint normal distribution
allows to model the conditional independence of ${\bf X}_t$ and $Y_{t}$ given $Z_t$.
Let us partition the conditional mean vector and covariance matrix of $(Y_t,\X_t)' \, | \, Z_t=k$
as follows:
\[ \mu_k=\left(  \begin{array}{c}
        \mu_{Y,k} \\
         \mu_{X,k} \\
  \end{array}
\right), \quad \quad
  \Sigma_k=\left(
  \begin{array}{cc}
    \Sigma_{Y,k} & \Sigma_{YX,k} \\
    \Sigma_{XY,k} & \Sigma_{X,k} \\
  \end{array}
\right), \quad k=1,\ldots,K. \]
Denote all the model parameters by $\Psi$. After estimating the joint distribution of CT and MRI sequences, a function for predicting the CT value is obtained by taking the conditional expectation of CT for given MRI sequences and parameters $\Psi$:
\[\{sCT\}_1^n=E[\{CT\}_1^n | \{MRI\}_1^n,\Psi],\]
where $sCT$ stands for pseudo-CT. Let us denote $Y^n=(Y_1,\ldots,Y_n)$. Since for any random variables $Y, X_1, X_2$, we have
$E(Y|X_1) =E_{X_2}[E(Y|X_1,X_2)|X_1]$,
we obtain:
\[ E[Y^n|\X^n,\Psi] =E_{Z^n}[E[Y^n|\X^n,Z^n,\Psi]|\X^n,\Psi].\]
For given $Z^n$, the observations are independent, and $E(Y_t|\X_t,Z_t,\Psi)$ can be calculated for each $t=1,\ldots,n$. In our case this is just a conditional expectation in a multivariate normal distribution:
\[ \tilde{\mu}_k(\x):=E(Y_t|\X_t=\x,Z_t=k,\Psi)=\mu_{Y,k}+\Sigma_{YX,k}\Sigma_{X,k}^{-1}(\x-\mu_{X,k}), \quad k=1,\ldots,K. \]
Thus, when we take the expectation over the distribution of $Z^n$, we obtain that the estimated CT image value for voxel $t$ is given by
\begin{equation} \label{substCT}
sCT_t= \sum_{k=1}^K P(Z_t=k|\X^n,\Psi) \tilde{\mu}_k(\X_t), \quad t=1,\ldots,n. \end{equation}
An important advantage of HMM compared to HMRF is that the
weights $P(Z_t=k|\X^n,\Psi)$ can be calculated exactly with the forward-backward algorithm.

\subsection{Hidden Markov random field models}

In the class of HMRF models, the observed variables depend also on latent variables indicating what class the corresponding voxel belongs to, whereas spatial dependence is accounted for through a MRF prior on the latent variables. Consider again the value of the CT image and MRI sequences at voxel $t$, that is $(Y_t,X_{t,1},\ldots,X_{t,m})$. Let $Z_t$ denote the hidden variable taking on one of the values $1,\ldots,K$. The joint distribution of the latent variables $Z=\{ Z_t \}$ can be formulated using the Gibbs field \citep{winkler}:
\[ p(z)=\frac{1}{W}\exp{\{-H(z)\}}, \quad W=\sum_z \exp{\{-H(z)\}}, \]
where $H(z)$ is the energy function defined as $H(z)=\sum_{c\in \mathcal{C}} V_c(z)$ and the potential $V$ depends on $z$ only through $c \in \mathcal{C}$, where $\mathcal{C}$ is the set of all cliques. As in the case of HMMs, we assume a first order neighbourhood structure for $Z$, which in $\mathbb{R}^3$ means that every voxel has six neighbours. Therefore, the possible cliques are singletons and neighbour pairs, and the potentials can be specified as
\[ V_{\{u\}} (z)= \alpha_k, \quad \mbox{if} \quad z_u=k,\]
\[ V_{\{u,v\}} (z)= \beta_k, \quad \mbox{if} \quad z_u=z_v=k.\]
\noindent The parameter vector $\alpha=\{ \alpha_1,\ldots,\alpha_K \}$ determines the prior probabilities for classes $1,\ldots,K$ and $\beta=\{\beta_1,\ldots,\beta_K \}$ determines the strength of spatial dependence. The case $\beta_1=\ldots=\beta_K=0$ corresponds to independent voxels and gives GMM. Thus, the interaction between the voxel classes is captured through the energy function. Observe that the considered parametrization is isotropic, that is the parameters $\beta_k$ do not depend on direction. The isotropy property justifies also using the Hilbert curve, which includes possible neighbour pairs in different directions with equal proportions. Note even that in the current parametrization, the voxel pairs with voxels belonging to different tissue classes do not contribute to the energy function.

We assume again that $(Y_t,\X_t)' \, | \, Z_t=k \, \sim \, \mathcal{N}(\mu_k,\Sigma_k)$, $k=1,\ldots,K$.
Therefore, the parameters to be estimated in HMRF in the case of normally distributed observations are $\Psi=(\alpha,\beta,\Theta)$, where $\Theta=((\mu_1,\Sigma_1),\ldots,(\mu_K,\Sigma_K))$.
The regression function obtained for calculating the CT estimate will have the same form as (\ref{substCT}). The important difference compared to the HMM case is that the weights $P(Z_t=k|\X^n,\Psi)$ cannot be computed analytically anymore. The weights have to be estimated using Gibbs sampling, see also \citet{David2014}.
\subsection{Parameter estimation}
The EM algorithm (Baum-Welch) and an algorithm following the ideas of the EM gradient algorithm by \citet{lange} is used to estimate the parameters of HMM and HMRF, respectively.
That the parameters of HMM can be estimated with the EM algorithm is another big advantage over HMRF, where the EM algorithm is not directly
applicable and should be combined with gradient methods. The estimation procedure by \citet{Hildeman2016} which is used for HMRF can be viewed as an EM algorithm, where the M-step is performed by one iteration of Newton's method.
The EM gradient algorithm contains several approximations, in
particular it maximizes the pseudo-likelihood instead of the likelihood. The algorithm is not fully developed yet.
Therefore, in the current work the models estimated with this algorithm are just used to obtain comparative numbers to HMM. We are not comparing HMRF models in terms of log-likelihood values, which in this application can also be very computer-intensive, since evaluating log-likelihood values requires Gibbs sampling.

Since both the EM algorithm and EM gradient algorithm can be very sensitive with respect to initial parameter values, we used a number of different initial parameter sets in the estimation process.
In the case of HMM, we used the parameter estimates for each single head as initial parameters. Thus, for every step of the leave-one-out cross-validation (LOOCV) scheme with 9 heads we estimated 9 models, the model with the highest log-likelihood value was chosen as the best model in each cross validation step. As convergence criterion the relative log-likelihood augmentation was used. In the case of HMRF, the initial parameter set was obtained as the one with highest log-likelihood value among a number of GMM models. As convergence criterion the norm of the difference between the consecutive parameter estimates was used. Mean absolute error was used to compare HMRF models. For the EM algorithm in estimating the GMM parameters, the initial parameter sets were chosen by using the $k$-means algorithm and agglomerative hierarchical clustering.
%
%
\subsection{Model assessment}
The main purpose of this study is to evaluate the class of HMM models in comparison to HMRF models in this particular application of generating pseudo-CT images. Observe that GMM can be seen as a special case of HMRF models and the results for this model class are presented to show how much accounting for the first order neighbourhood structure helps to improve the model. The loss function we use for measuring errors between CT and pseudo-CT images is the voxel-wise absolute error. Thus, for each CT estimate its mean absolute error (MAE) is calculated and this is one of the main criteria for measuring goodness of estimated CT images. Let the number of measurement voxels for head $l$ be $n_l$, then the mean absolute error for head $l$ is given by
   \[ \mbox{MAE}_l =\frac{1}{n_l} \sum_{j=1}^{n_l} |CT_{l,j}-sCT_{l,j}|.\]
For studying model behaviour in different regions of the head, we have used smoothed residual plots. Smoothed residuals have been calculated by moving over the CT range with non-overlapping windows of size 20, for each window the average of the residuals (or their absolute values) is computed. Smoothed residual plots enable to observe the general behaviour of the residuals for these models and to point out areas where the three models differ at most.

The complexity of the models increases with increasing number of underlying states (number of tissue classes). To obtain a fair comparison between the three model classes, it is essential to study how the number of states, the number of patients used for training a model and variability between patients affects modeling. To investigate this, we have run the LOOCV scheme for the nine patients with 5 and 8 tissue classes, for HMM also with 10 tissue classes. Since the first modeling round demonstrated that the fitted models give bad results for some heads, we have run the LOOCV scheme also for a subset of five heads only.

\section{Model comparisons} \label{mudvordlus}

\subsection{Modeling results for 9 patients}
In Table \ref{MAEsummary9}, a summary of MAEs is given for HMM, HMRF and GMM for different number of underlying state classes. To clarify the comparison, the average of MAEs over the nine heads is presented in the last row of the table.  The MAE value for each head is calculated using the model where the respective head was excluded when training the model parameters. Comparing the rows for different heads in this table shows directly that none of the three model classes seems to work for heads 3, 5, and 7, head 6 is on the borderline. The MAE values for these heads are much larger in comparison to heads 1, 2, 4, 8 and 9 and increasing the number of underlying tissue classes does not give any improvement, either. The best result is obtained for HMM with $K=8$. Increasing the number of underlying states to $K=10$ improves MAEs for `good' heads only slightly, we shall comment more on this issue in Section \ref{arutelu}. We can see that HMM and GMM give very similar results for $K=5$. Increasing the number of states from 5 to 8 does not improve MAE values for GMM.  For HMRF with $K=8$ we experienced numerical difficulties in estimating the models. Since it is clear from Table \ref{MAEsummary9} that the nine heads cannot be treated together, we left two cells for HMRF with $K=8$ empty.
\begin{table}[th]
\caption{MAEs for HMM, HMRF and GMM with different number of states. \label{MAEsummary9}}
\centering
\begin{tabular}{c|ccc|cc|cc}
\hline\hline
       &\multicolumn{3}{c}{HMM}       & \multicolumn{2}{|c|}{HMRF}    & \multicolumn{2}{c}{GMM}   \\
Head   &    K=5  & K=8  & K=10  & K=5  & K=8   & K=5    & K=8   \\
\hline
  1    & 160.21  & 146.31  & 144.38 & 149.42 & 203.94 & 161.24 & 169.96 \\
  2    & 170.83  & 146.15  & 150.11 & 157.66 & 177.91 & 175.45 & 166.92 \\
  3    & 293.24  & 297.35  & 298.28 & 291.86 & 302.68 & 291.48 & 298.37 \\
  4    & 190.20  & 157.00  & 154.78 & 173.47 & 177.48 & 194.79 & 186.38 \\
  5    & 251.54  & 259.67  & 271.97 & 256.20 & 302.37 & 256.51 & 267.80 \\
  6    & 211.59  & 199.34  & 221.07 & 198.01 & 238.29 & 215.09 & 224.79 \\
  7    & 355.96  & 351.73  & 350.19 & 368.19 &  & 347.40 & 341.20 \\
  8    & 183.09  & 153.21  & 149.71 & 167.52 & 162.44 & 181.53 & 171.75 \\
  9    & 170.33  & 153.87  & 151.88 & 161.78 &  & 171.26 & 167.24 \\
\hline
  Mean & 220.78  & \textbf{207.18} & 210.26 & 213.79 &  & 221.64 & 221.60 \\
\hline\hline
\end{tabular}
\end{table}

\begin{figure}[ht]
\centering
\includegraphics[scale=0.5]{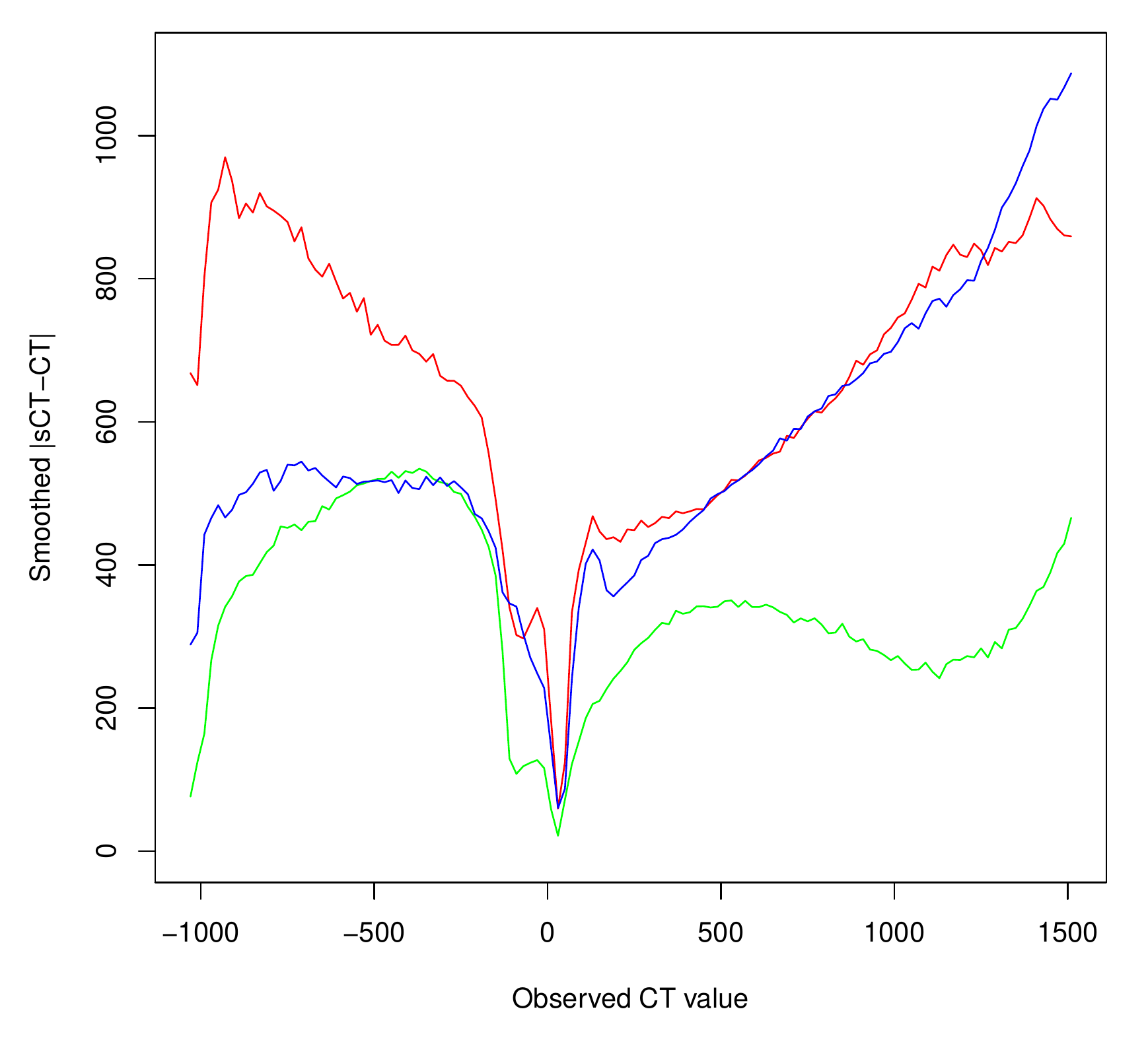}
\caption{Smoothed absolute values of the residuals for the full HMM model with 8 states: head 1 - green, head 3 - blue, head 7 - red. \label{GoodBad}}
\end{figure}

Since MAE is very summarizing as a measure of goodness, we have also calculated and compared smoothed residuals and absolute values of the smoothed residuals for the studied models. Smoothed residuals have always been computed for non-overlapping windows of size 20, meaning that the average residual value in each window of this size is calculated and presented. In Figure \ref{GoodBad} we have plotted the smoothed absolute values of residuals corresponding to HMM with 8 state classes for three heads: one `good' head (head 1) and two `bad' heads (head 3 and head 7). The green line corresponding to head 1 represents the typical residual behaviour for these models, the same pattern can be seen for example in \citet{adam2}. The blue and red line on the other hand illustrate that the model does not work at all for head 3 and head 7. Thus, Figure \ref{GoodBad} is a warning signal, because if we want to use these latent variable models in practice, we want them to be robust in regard to different heads.
\begin{table}[h!]
\centering
\caption{MAEs for HMM ($K=8$) from the LOOCV scheme, where $[-i]$ stands for the model with head $i$ being excluded from the training set. \label{hmm8mae}}
\begin{tabular}{c|ccccccccc}
\hline\hline
 Head      &	[-1] &	[-2] & [-3] & [-4] & [-5] & [-6] & [-7] & [-8] & [-9] \\
\hline
  1
       & 146.31 & 144.96 & 142.37 & 147.48 & 135.68 & 139.04 & 141.91 & 147.74 & 146.17 \\
  2
       & 145.80 & 146.15 & 142.64 & 147.94 & 137.89 & 141.56 & 142.49 & 147.73 & 146.28 \\
  3
       & 293.46 & 292.65 & 297.35 & 292.05 & 290.75 & 289.85 & 299.26 & 292.95 & 293.19 \\
  4
       & 151.98 & 151.91 & 148.38 & 157.00 & 145.70 & 149.41 & 146.56 & 155.58 & 153.46 \\
  5
       & 218.76 & 219.27 & 217.91 & 217.49 & 259.67 & 239.26 & 222.24 & 217.46 & 218.07 \\
  6
       & 183.24 & 183.34 & 181.24 & 184.24 & 208.66 & 199.34 & 182.44 & 183.63 & 183.44 \\
  7
       & 318.16 & 318.67 & 323.02 & 316.76 & 320.35 & 314.42 & 351.73 & 316.44 & 317.14 \\
  8
       & 149.32 & 148.52 & 146.09 & 151.98 & 137.93 & 142.30 & 143.53 & 153.21 & 149.67 \\
  9
       & 152.88 & 151.79 & 149.18 & 154.76 & 145.60 & 147.65 & 149.32 & 154.89 & 153.87 \\
\hline\hline
\end{tabular}
\end{table}

Table \ref{hmm8mae} presents MAE values for the LOOCV scheme in the case of HMM with 8 state classes. In each row we can see the MAEs for the respective head for the nine estimated models (each model is trained with 8 heads). The table illustrates that LOOCV does not have any particular effect on parameter estimation, the row-wise MAE values are stable. In particular, the table shows that among the nine heads we cannot point out any single outlier. On the other hand, this does not mean that all the heads are forming a homogeneous group. Table \ref{hmm8mae} suggests that there is a homogeneous subset of `good' heads (1, 2, 4, 8, 9) and the rest of the heads seem not to fit into this subset.

Based on these preliminary numerical results, we continued modeling using the subset of `good' heads (1, 2, 4, 8, 9) only. Considering the heads that behave homogeneously should allow us to get better comparisons between HMM, HMRF and GMM in this particular application.
\subsection{Modeling results for the subset of 5 patients}
A summary of MAE values for the subset models is presented in Table \ref{MAEsummary5}. Because of numerical problems when estimating the parameters for 8 state classes, three models in Table \ref{MAEsummary5}
(one for GMM and two for HMRF) were estimated by adding some noise to the data. These MAE values are marked with $^\ast$. The best result in terms of MAE is received for HMM, the models with $K=5$ and $K=8$ give practically the same MAEs. For GMM increasing the number of tissue classes from 5 to 8 improves the summary measure of goodness by approximately 10 units. The MAE values in Table \ref{MAEsummary5} show that the HMM models behave better than the HMRF models, the best average MAEs are 142 and 149, respectively. In terms of MAE, the performance of GMM with $K=8$ is similar to the performance of HMRF with $K=8$.
\begin{table}[th]
\centering
\caption{MAEs for the subset models with different number of states. \label{MAEsummary5}}
\begin{tabular}{c|cc|cc|cc}
\hline\hline
       &\multicolumn{2}{c}{HMM}       & \multicolumn{2}{|c|}{HMRF}    & \multicolumn{2}{c}{GMM}   \\
Head   &    K=5  & K=8  &  K=5  & K=8   & K=5    & K=8   \\
\hline
  1    & 133.57  & 133.86  & 137.80 & 158.51 & 153.69 & 144.92  \\
  2    & 138.69  & 141.30  & 137.47 & 158.98 & 160.12 & 153.39$^\ast$  \\
  4    & 154.97  & 149.47  & 159.14 & 162.37 & 179.56 & 166.32  \\
  8    & 139.34  & 136.32  & 145.09 & 146.71$^\ast$ & 160.55 & 150.91  \\
  9    & 143.90  & 151.02  & 165.41 & 152.39$^\ast$ & 169.99 & 160.68  \\

\hline
  Mean & \textbf{142.09}  & \textbf{142.39}  & 148.98 & {155.79} &  164.78  & 155.24 \\
\hline\hline
\end{tabular}
\end{table}

\begin{figure}[th]
\centering
\includegraphics[scale=0.5]{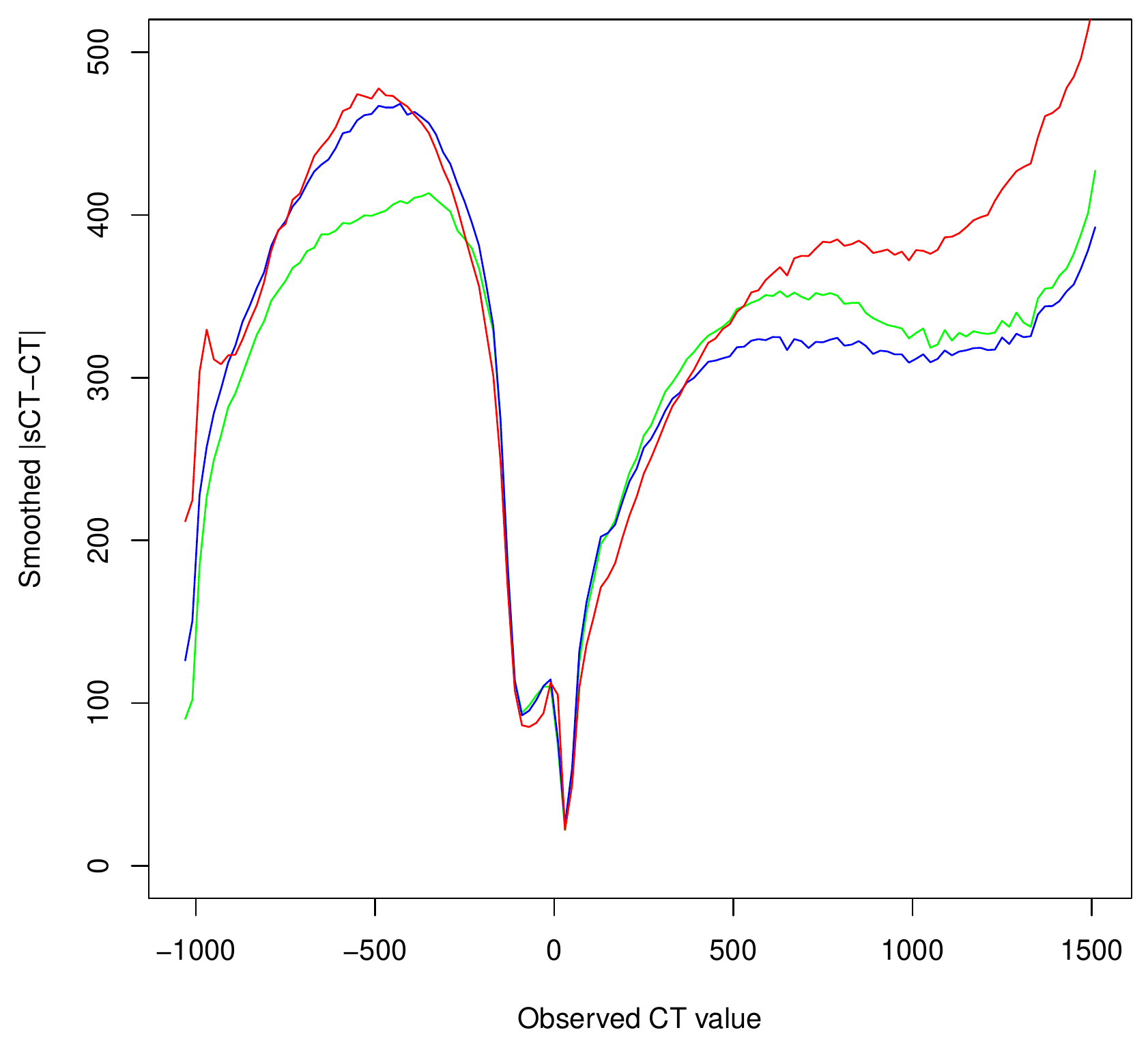}
\caption{Smoothed absolute residuals for the subset models with 5 classes. The models: HMM - green, HMRF - blue, GMM - red. \label{absres}}
\end{figure}

\begin{figure}[th]
\centering
\includegraphics[scale=0.5]{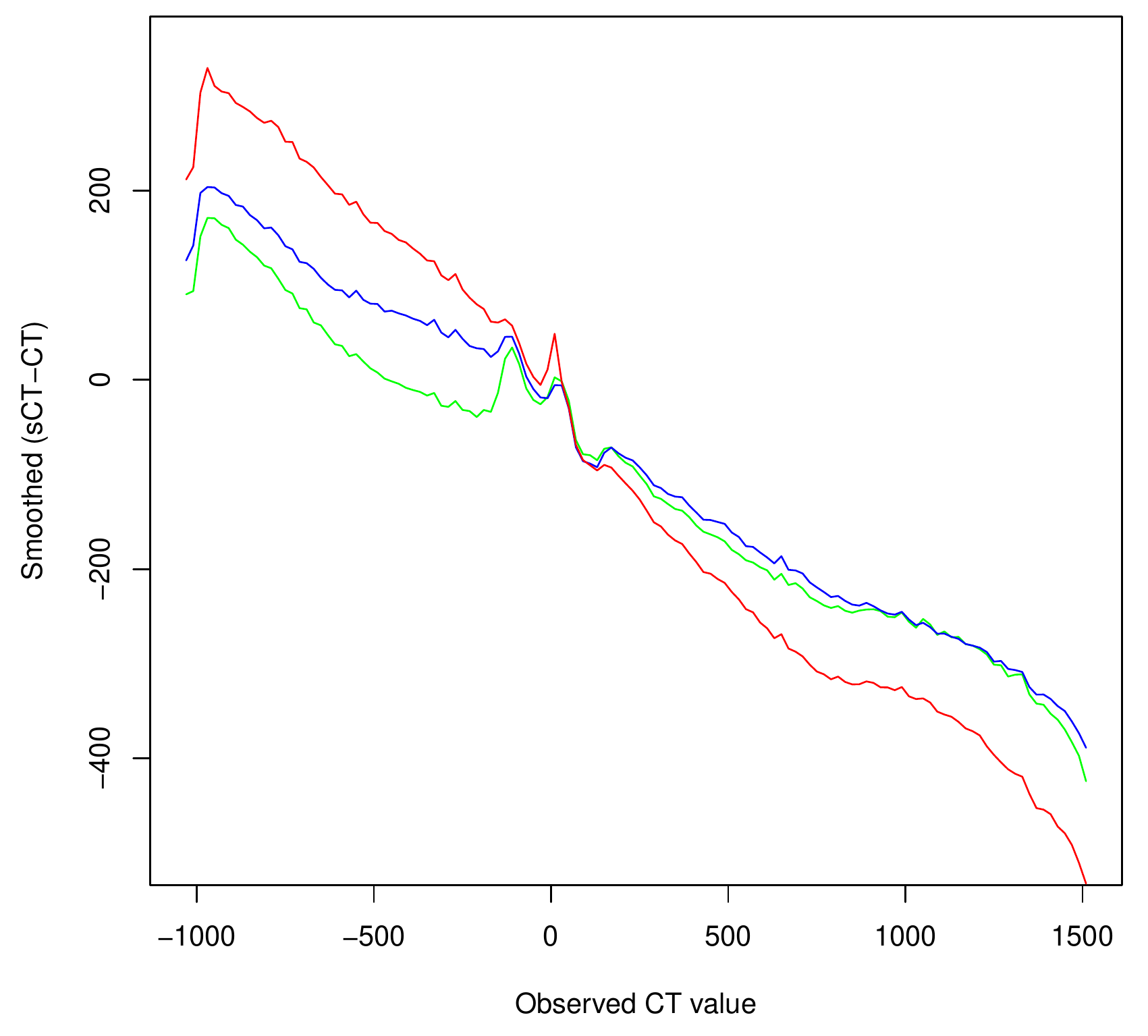}
\caption{Smoothed residuals for the subset models with 5 classes. The models: HMM - green, HMRF - blue, GMM - red. \label{justres}}
 \end{figure}

\begin{figure}[th]
\centering
\includegraphics[scale=0.5]{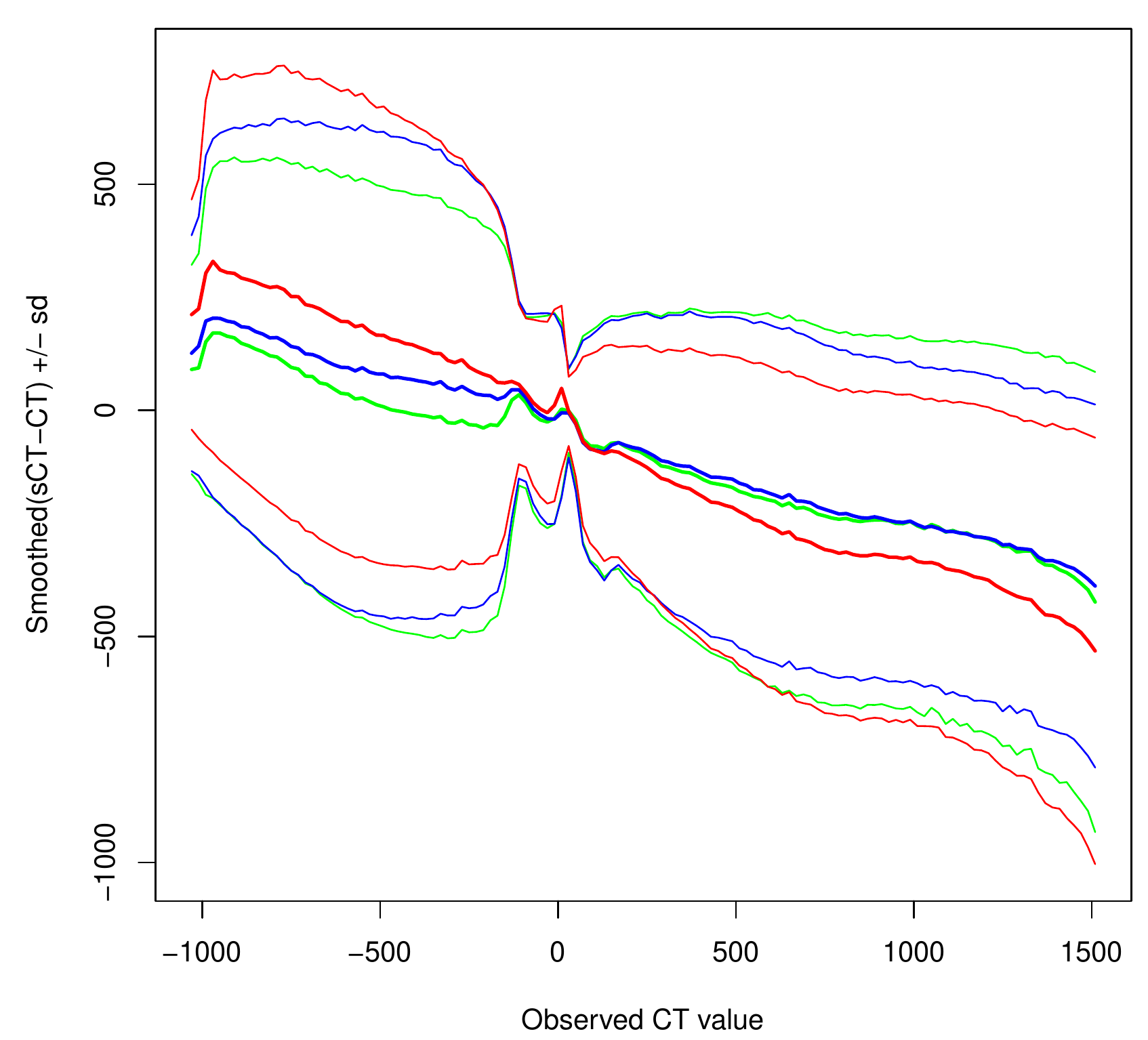}
\caption{Smoothed residuals for the subset models with 5 classes together with $\pm$ standard deviation lines. The models: HMM - green, HMRF - blue, GMM - red. \label{reswithsd}}
\end{figure}

Again, since MAE is very summarizing as a measure of goodness, we also present the smoothed residuals and the absolute values of the smoothed residuals for the subset models with 5 state classes. In Figure \ref{justres}, the smoothed residuals for window size 20 are plotted, meaning that the average residual value of each window is calculated and plotted. The average is calculated over all the five heads. In Figure \ref{absres}, the same is done for the absolute values of the residuals. Figure \ref{absres} shows that neither HMM or HMRF is superior over the whole CT observation range: on average, the HMM model gives better result than HMRF for the negative CT values, whereas HMRF has slightly lower absolute residuals for the positive CT values. Figure \ref{justres} demonstrates that all the models tend to overestimate the true negative CT values and underestimate the true positive CT values. In Figure \ref{reswithsd}, the smoothed residuals are plotted together with $\pm$ standard deviation values for the subset models with 5 classes. This figure illustrates that variation in the residuals is huge.

\section{Discussion and conclusions} \label{arutelu}
One of the main aims of this study was to compare the performance of HMMs and HMRF models in estimation of CT images. Since HMM is computationally much faster and easier to handle (parameters can be estimated with the EM algorithm and weights in the regression function can be computed exactly with the forward-backward probabilities), this model class has a clear advantage over HMRF in applications with big data amounts if model diagnostics are comparable for both models. One big advantage of HMM models when applying the ML method for estimating the parameters is that the log-likelihood value can be calculated analytically. This enables to employ the ML approach fully. The log-likelihood values can be used in model comparisons and are valuable information on how the chosen modeling approach works, since we are able to calculate different information criteria. In the case of HMRF models log-likelihood values can be evaluated by Gibbs sampling, but this can be very computer-intensive, because sampling from random field distributions for given parameter sets can require many iterations due to poor mixing of the Gibbs sampler.

Our results confirm that model diagnostics are better for HMM than for HMRF in this particular application. Comparison of MAEs shows that  HMM performs better than HMRF (see the results for HMM and HMRF in Table \ref{MAEsummary5}). Concerning residual behaviour, Figures \ref{absres} and \ref{justres} show that neither HMM or HMRF is superior in the whole CT region. In Tables \ref{meantable5} and \ref{meantable8} we have presented the estimates of expected CT values $\mu_{Y,k}$ for our subset models with $5$ and $8$ states, respectively.
%
%
\begin{table} [h!]
\centering
\caption{Estimates of expected CT values $\mu_{Y,k}$ for the subset models with $K=5$. \label{meantable5}}
\begin{tabular}{c|rrrrr}
 \hline\hline
   HMM  &  &  &  &  &    \\
  \hline
  M1  & -1018 & -527 & -13 & 32 & 654  \\
  M2 & -1018 & -512 & -3 & 31 & 657    \\
  M3  & -1020 & -590 & -12 & 31 & 500   \\
  M4  & -1018 & -518 & -10 & 31 & 600 \\
  M5  & -1018 & -515 & -12 & 32 & 652 \\
 \hline
  HMRF  &  &  &  &  &    \\
 \hline
 M1 & -1019 & -590 & -17 & 32 & 546 \\
 M2 & -1019 & -544 & -9 & 31 & 599 \\
 M3 & -1021 & -617 & -18 & 31 & 475 \\
 M4 & -1020 & -585 & -15 & 31 & 485 \\
 M5 & -1021 & -608 & -17 & 32 & 513 \\
 \hline
  GMM  &  &  &  &  &    \\
 \hline
 M1 & -1021 & -649 & -7 & 32 & 499 \\
 M2 & -1024 & -748 & 2 & 32 & 471 \\
 M3 & -1024 & -743 & -10 & 31 & 439 \\
 M4 & -1024 & -757 & -4 & 31 & 429 \\
 M5 & -1024 & -755 & -7 & 32 & 473 \\
 \hline\hline
\end{tabular}
\end{table}
With M1,$\ldots$,M5 we denote the best models when heads $1,\ldots,5$, respectively, were excluded when training a model. Figure \ref{absres} together with Table \ref{meantable5} suggests that residual behaviour is mostly determined by the CT group means $\hat{\mu}_{Y,k}$.  Tables \ref{meantable5} and \ref{meantable8} show that increasing the number of tissue classes $K$ might not help in obtaining a more uniform distribution of CT group means over the whole CT range. This might also explain why in the case of nine heads, MAE for HMM with $K=8$ is slightly better than with $K=10$. To guarantee a more uniform location of the CT group means, one should maybe fix a certain number of CT group means and estimate the models under restrictions. This indicates that purely model-based approach where everything is estimated from data (unsupervised modeling) might not be justified in this application and possibilities for including appropriate available information to HMM in the best way (supervised modeling) should be investigated in the future. One possible direction could be combining regression with segmentation and atlas-based approach. With HMMs it is easy to perform segmentation and this can be done in different ways \citep{intech}.
In the current application, when underlying states have physical meaning (tissue classes), it is realistic to assume that in some regions of the head the underlying states can be revealed. This basically means that a certain amount of states can be assumed to be known, and as the study in \citet{peep} shows, even the tiny fraction of truth can make a big improvement in inferences.

An important advantage of HMM in comparison to HMRF and GMM is its stability.  In Tables \ref{meantable5} and \ref{meantable8}, for all the five HMM models the number of positive and negative CT group means is the same and the means do not differ so much between the models. In the case of HMRF and GMM, for $K=8$ location of group means varies over M1--M5. Besides, our computations show that HMRF is sensitive with regard to initial values and small changes in the data (for example adding some noise). That HMM is more robust than HMRF has been demonstrated also in other studies \citep{Fjortoft2003}.

Modeling results for 9 and 5 heads illustrate the sensitivity of the considered models with respect to data. It is worrying that MAE can differ so much depending on a head. Previous studies \citep{adam1, adam2, adam3} for the same application with GMM have not reported the robustness problem. It is essential to investigate this issue and find out why do the models not work for some heads and what characterizes those heads, because this might determine the potential of the whole approach in practice.

We can conclude that both HMM and HMRF give better results than GMM, meaning that including the spatial dependence information improves the model. The comparison of HMM and HMRF shows that HMM has definitely more advantages. Therefore, as HMMs have better performance than HMRF models
in the current application, they deserve a further study for investigating their
potential in obtaining good estimates of CT images.

\newpage

\begin{table} [h!]
\centering
\caption{Estimates of expected CT values $\mu_{Y,k}$ for the subset models with $K=8$. \label{meantable8}}
\begin{tabular}{c|rrrrrrrr}
 \hline\hline
   HMM  &  &  &  &  & & &  &     \\
  \hline
 M1 & -1021 & -659 & -47 & -39 & 29 & 34 & 43 & 914 \\
 M2 & -1021 & -655 & -31 & -6 & 28 & 33 & 44 & 922 \\
 M3 & -1022 & -660 & -47 & -41 & 28 & 32 & 41 & 890 \\
 M4 & -1021 & -657 & -55 & -40 & 28 & 33 & 44 & 868 \\
 M5 & -1022 & -672 & -48 & -41 & 29 & 34 & 43 & 924 \\
 \hline
   HMRF  &  &  &  &  & & & &     \\
 \hline
 M1 & -1021 & -645 & -68 & -42 & 28 & 34 & 37 & 940 \\
 M2 & -1022 & -678 & -35 & -22 & 28 & 33 & 37 & 936 \\
 M3 & -1022 & -655 & -52 & -43 & 27 & 32 & 37 & 918 \\
 M4$^\ast$ & -1023 & -1009 & -504 & -142 & -24 & 33 & 34 & 763 \\
 M5$^\ast$ & -1024 & -1009 & -520 & -62 & -27 & 33 & 34 & 942 \\
 \hline
   GMM  &  &  &  &  & & & &     \\
  \hline
 M1 & -1024 & -908 & -55 & -15 & 32 & 34 & 212 & 1129 \\
 M2$^\ast$ & -1024 & -780 & -238 & -9 & 31 & 33 & 436 & 747 \\
 M3 & -1024 & -781 & -24 & -19 & 31 & 33 & 241 & 1097 \\
 M4 & -1024 & -752 & -36 & -2 & 26 & 33 & 50 & 800 \\
 M5 & -1024 & -752 & -36 & 12 & 28 & 34 & 49 & 875 \\
 \hline\hline
\end{tabular}
\end{table}

\subsection*{Acknowledgments}
This work is supported by the Swedish Research Council grant (Reg.No.~340-2013-5342) and Estonian institutional research funding IUT34-5. Adam Johansson is acknowledged for providing us with data.
\bibliographystyle{apa}
\bibliography{substCT}

\end{document}